\documentclass[article,11pt]{revtex4-1}

\usepackage{amsmath}    
\usepackage{graphicx}   
\usepackage{hyperref}   
\pdfoutput=1
\begin{document}

{\Large\centering
\textbf{Micro--structured crystalline resonators for optical frequency comb generation} \\ }
{\centering
\textit{
I. S. Grudinin,$^{*}$ and Nan Yu\\
Jet Propulsion Laboratory, California Institute of Technology, \\ 4800 Oak Grove dr., Pasadena, CA 91109, USA \\ $^*$Corresponding author: grudinin@jpl.nasa.gov} \par
}

\textbf{
Optical frequency combs have recently been demonstrated in micro--resonators through nonlinear Kerr processes\cite{sciencereview}. Investigations in the past few years provided better understanding of micro--combs and showed that spectral span and mode locking are governed by cavity spectrum and dispersion\cite{scalinglaws,gaetaroute,grudininCrossing,herrCrossing}. While various cavities provide unique advantages\cite{herrCrossing, grudininCrossing, silica,vahalacombs}, dispersion engineering has been reported only for planar waveguides\cite{gaetaCMOS}. In this Letter, we report a resonator design that combines dispersion control, mode crossing free spectrum, and ultra--high quality factor. We experimentally show that as the dispersion of a MgF$_2$ resonator is flattened, the comb span increases reaching 700 nm with as low as 60 mW pump power at 1560 nm wavelength, corresponding to nearly 2000 lines separated by 46 GHz. The new resonator design may enable efficient low repetition rate coherent octave spanning frequency combs without the need for external broadening, ideal for applications in optical frequency synthesis, metrology, spectroscopy, and communications. 
}

First demonstrations of Kerr optical frequency combs in micro--resonators\cite{firstcomb,grudinin-caf2,savchenkov-caf2} have stimulated great interest and intense research activity. Rapid progress has been highlighted by such accomplishments as octave spanning comb spectrum\cite{silicaoctave, nitrideoctave}, soliton formation and mode locking \cite{solitonsMgF2, purdue, oewaves74fs}. Practical use of micro--combs also begins to emerge in timekeeping \cite{oewavesClock,diddamsClock} and communications \cite{terabit}. While external broadening can be used to extend the spectral span to an octave\cite{solitonsMgF2} necessary for self--referencing, direct generation of octave--spanning combs with measurable repetition rates in a single device remains highly desirable. Latest research results show that the comb span is primarily limited by the wavelength dependence of cavity dispersion \cite{scalinglaws,gaetaroute, herrCrossing, grudininCrossing}. Moreover, most resonators used for comb generation support more than one family of modes, leading to unavoidable mode crossings that can prevent soliton formation \cite{herrCrossing, grudininCrossing}. Thus a high quality factor (Q) resonator supporting a single mode family with properly engineered dispersion may be necessary for efficient direct generation of a coherent octave spanning comb.

It is well known that micro--structuring of fibers and waveguides can provide control over the wavelength dependence of dispersion and make it possible to achieve anomalous geometric (waveguide) dispersion for improved supercontinuum and comb generation\cite{supercontinuum}.
Similarly, dispersion engineering is possible in silicon nitride planar waveguide resonators\cite{gaetaCMOS}. Nevertheless, obtaining anomalous GVD in a broad range of wavelengths remains challenging. On the other hand, the whispering gallery mode (WGM) resonators have demonstrated extremely high quality factors\cite{grudininMicro, vahalacombs} and can be easily made from large selection of nonlinear optical materials\cite{WGMreview}. The high quality factor of crystalline WGM resonators leads to high efficiency of nonlinear optical processes. However, dispersion modification in WGM resonators has only been limited to moderate mostly wavelength independent shifts towards normal dispersion\cite{ilchenkoComp, silicaoctave}. The total cavity group velocity dispersion (GVD) consists of the material and geometric (waveguide) contributions. While groups of modes with effectively anomalous geometric dispersion can be found\cite{selectable}, the geometric dispersion in a WGM resonators is generally normal\cite{dispersion}. This has also spurred interest in normal GVD comb regime\cite{oewaves74fs, purdue}.
Regardless of the dispersion sign, engineering its wavelength dependence will be critical for efficient wide band comb generation.

\begin{figure}[htbp]
\centering
\includegraphics[width=12 cm]{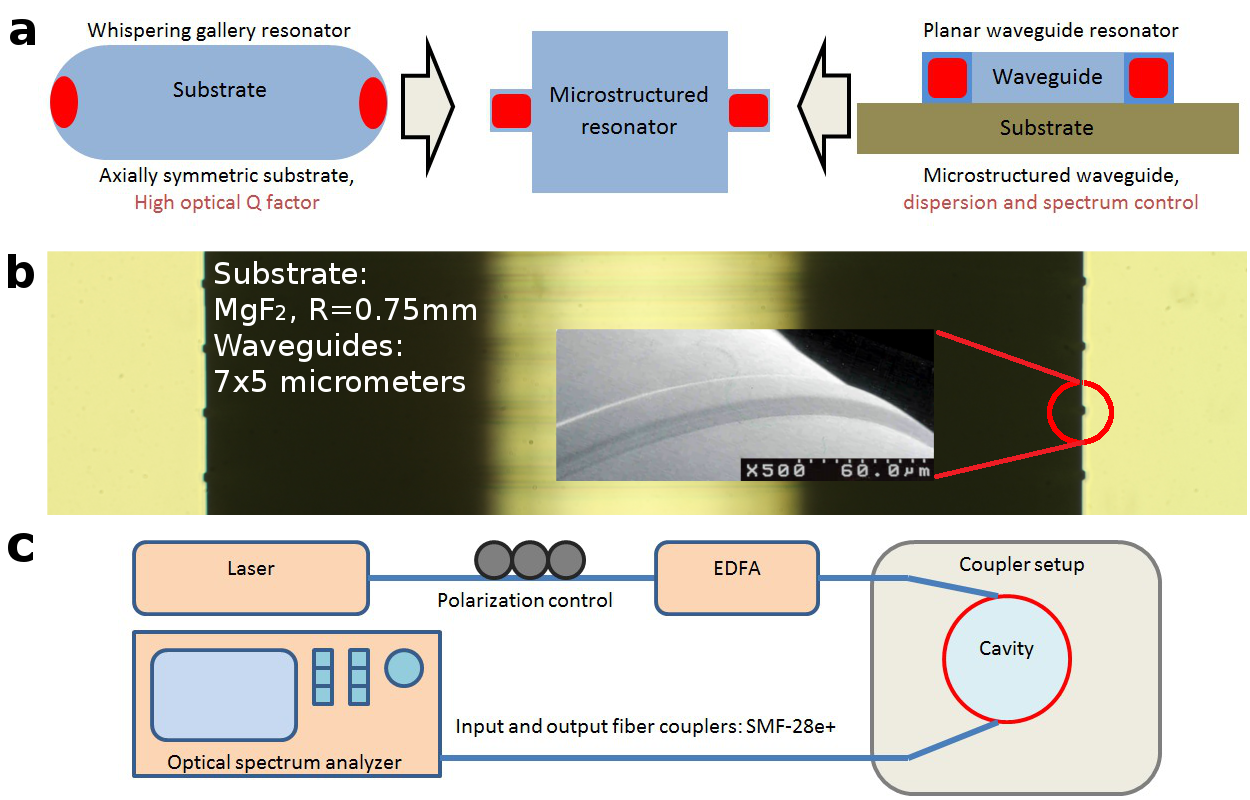}
\caption{\textbf{Microstructured cavity configuration and experimental setup.} \textbf{a}, The cavities share the axial symmetry of substrate with WGM resonators and a microstructured waveguide with planar ring cavities. \textbf{b}, Example of four, batch fabricated microstructured waveguides on a MgF$_2$ substrate (shadow photography) with an SEM photo inset. c) Schematics of the experimental setup.}
\label{fig:cavity}
\end{figure}

We here demonstrate a new resonator design which combines the dispersion control similarly to  microstructured waveguides with the high Q factor and material versatility available to WGM resonators as shown in Fig. \ref{fig:cavity}. The resonator consists of a substrate and a micro--structured waveguide sharing the same axial symmetry. The supported modes, their quality factors and dispersion are defined by the waveguide shape and its relation with the substrate, providing the needed design flexibility. We show by numerical simulation, that dispersion control through waveguide shaping works well in the presence of a substrate. Moreover, the interaction of the substrate with the shaped waveguide allows controlling the families of supported modes and their quality factors similarly to WGM resonators\cite{singlemode,singlemodeQ}. As a demonstration, we fabricated a number of microstructured resonators and were able to show that as the calculated cavity GVD is gradually flattened the comb span increases significantly, reaching over 700 nm around the 1560 nm pump with only 60 $\mu$W of coupled power. 

We design our resonator with the rectangular shape of the waveguide. Other shapes can be used for fine--tuning the dispersion as is shown in Supplementary Information. For the target resonator free spectral range (FSR) $F=46$ GHz around the pump wavelength of $\lambda=1.56$ $\mu$m the radius of the cavity is approximately given by $R\simeq c/(2\pi n_e F)\simeq 750$ $\mu$m, where $n_e\simeq 1.38$ is the extraordinary refractive index and c is the speed of light in m/s. We limit the spectrum of our cavity to only one family of TE (electric field along the cavity axis) modes by enforcing the geometric condition developed for the WGM resonators\cite{singlemode}: $1>1.565w^2h/(750-h)>1/4$, where h is the height of the waveguide and w --- its width in micrometers. For $h=5$ $\mu$m the condition holds if $4.9<w<9.7$, leaving us enough flexibility to change waveguide geometry while staying within the single mode limit.

In order to find the dependence of dispersion on wavelength we compute the mode frequencies numerically. The total cavity GVD includes material and waveguide contributions and is related to mode frequencies\cite{dispersion}:
\begin{equation}
D=\frac{cD_2}{2\pi\lambda^2RF^3}\times 10^6\left[\frac{ps}{nm\cdot km}\right], \nonumber
\end{equation}
where  $F=0.5(f_{l+1}-f_{l-1})$ is FSR near $\lambda$, $D_2=f_{l+1}-2f_l+f_{l-1}$, and $f_l$ is the optical mode frequency in GHz corresponding to $l$ field maxima along the waveguide ring. By modelling a number of shapes we found that a rectangular waveguide with height of 5 $\mu$m and width of 7 $\mu$m will have a flattened and reduced overall dispersion around the 1560 nm pump wavelength which should lead to broadened comb span\cite{scalinglaws}. We also found that the flattened band of dispersion in a micro--structured resonator can approach one octave (Supplementary Information).
\begin{figure}[htbp]
\centering
\includegraphics[width=10 cm]{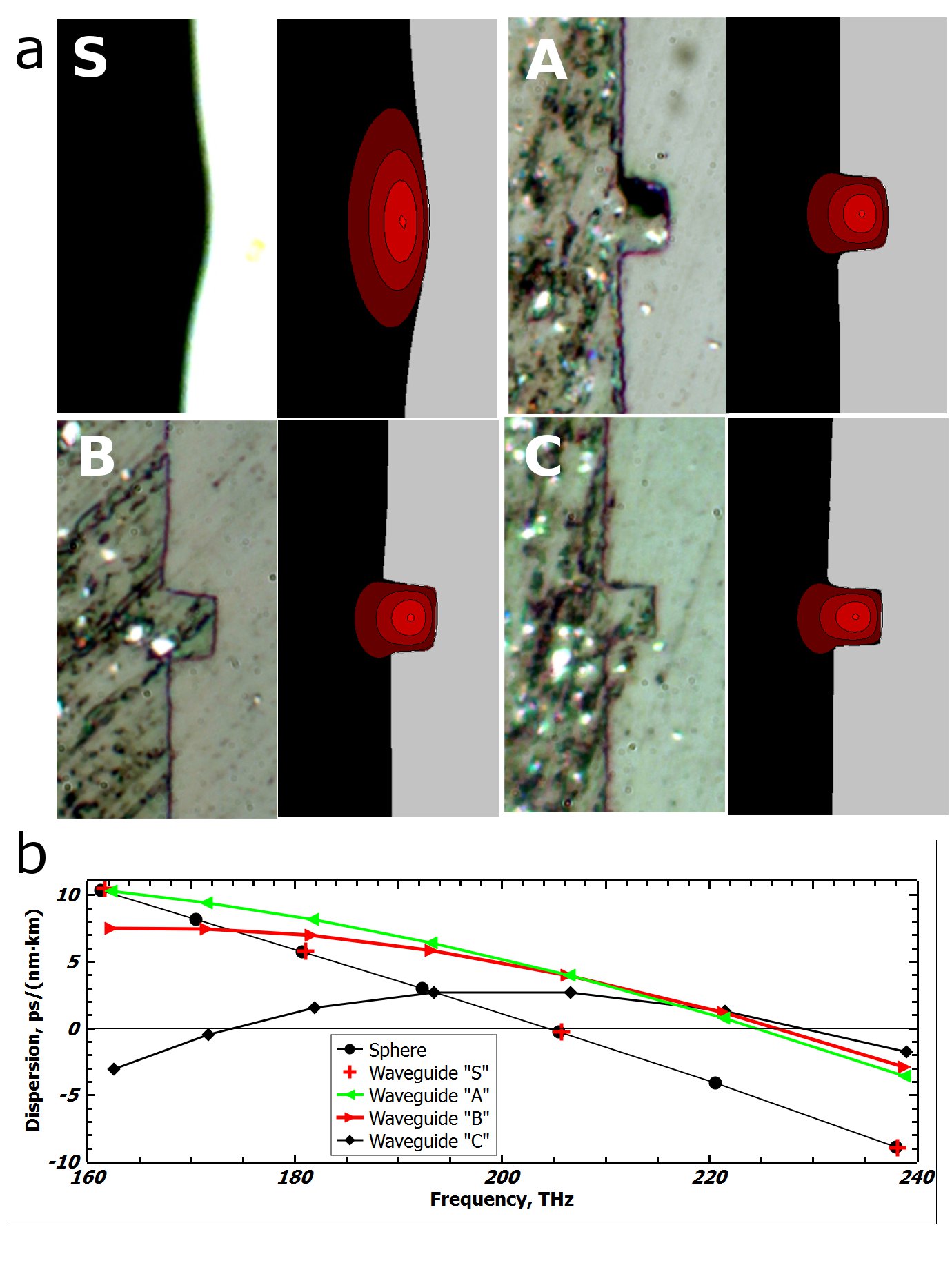}
\caption{\textbf{Microstructured waveguides and corresponding dispersion.} \textbf{a}, Each of the 8 images represents an area sized $25\times 45$ micrometers. The optical images of the waveguide cross sections are shown along with the mode intensity maps obtained with FEM modelling. \textbf{b}, Numerically computed total cavity dispersion for the waveguides shown in a). The waveguide ``S'' with Gaussian waveguide shape has the same dispersion as an ideal sphere.}
\label{fig:waveguides}
\end{figure}

We have fabricated several resonators having the same radius of 750 $\mu$m but different shapes of the waveguide. By using a non--destructive imaging technique developed in our lab we obtained the cross sections of the waveguides. These images were then converted into meshed geometries for our FEM modelling tool which provided the wavelength dependence of GVD as shown in Fig. \ref{fig:waveguides}.
\begin{figure}[htbp]
\centering
\includegraphics[width=16 cm]{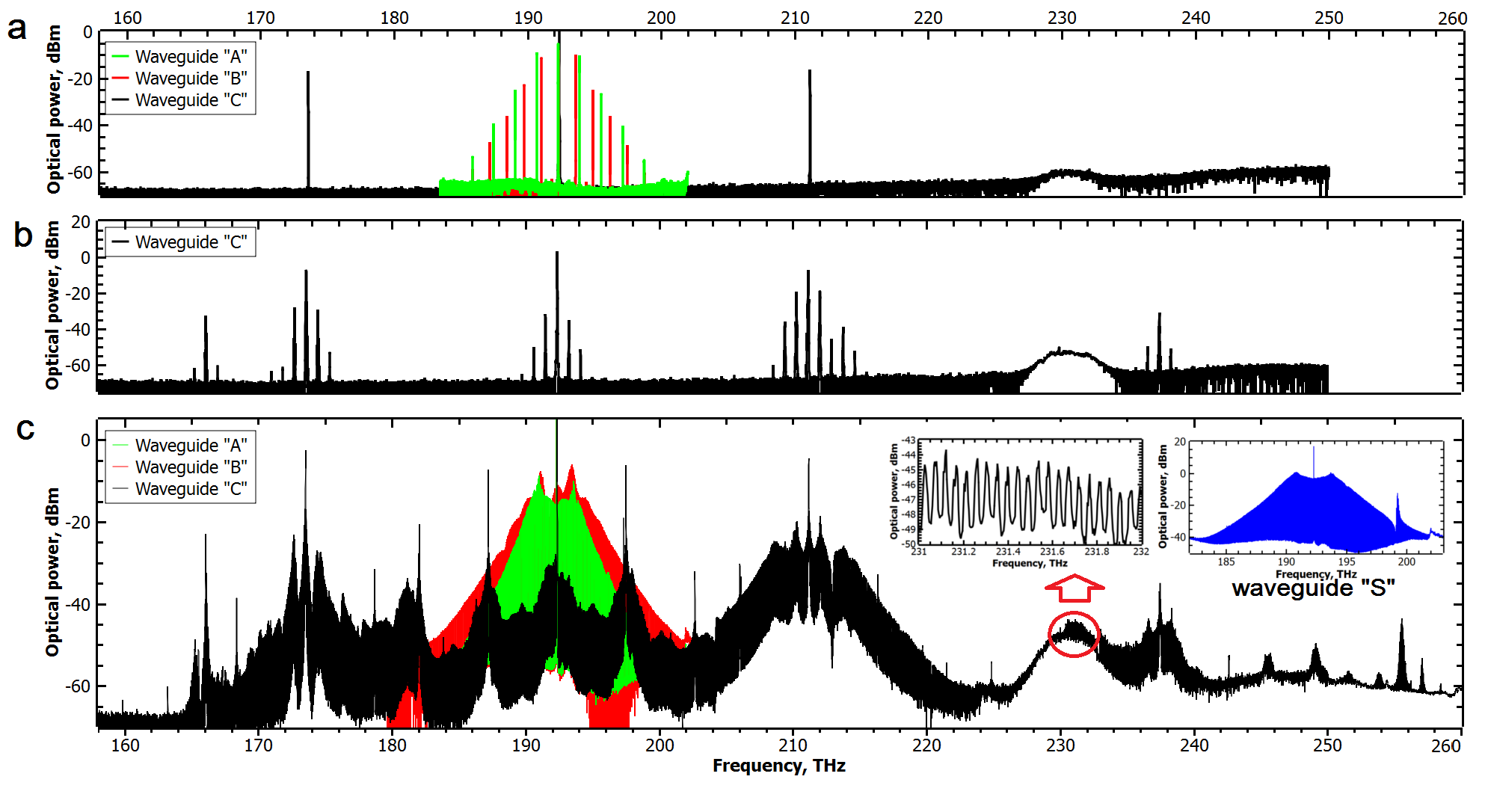}
\caption{\textbf{Frequency combs generated in microstructured resonators with 300 mW of pump at $\lambda=$1560 nm (192.4 THz).} \textbf{a}, The primary comb in waveguide ``C'' starts at N=408 in contrast to N$\simeq$30 in waveguides ``A'' and ``B''. \textbf{b}, Secondary comb formation in waveguide ``C'' starts as the laser detuning is reduced. \textbf{c}, Comb states at a minimum stable detuning. The comb from the waveguide ``C'' contains nearly 2000 lines spanning 100 THz. Insets show cavity FSR--spaced comb lines. The comb from the waveguide ``S'' shows evidence of avoided mode crossing\cite{herrCrossing, grudininCrossing}. }
\label{fig:combs}
\end{figure}
We pumped the blue side of cavity resonance and relied on a soft thermal lock self--stabilization technique to excite the combs. The pump power was fixed at 300 mW, while the actual power coupled to the resonator is given by the coupling efficiency. The primary and the final comb states produced by the micro--structured resonators are presented in Fig. \ref{fig:combs}. The resonator and coupling parameters are summarized in Table \ref{tab:resonators}. As the detuning was reduced we observed deviations from previously reported comb formation dynamics\cite{yu3,universal} (Supplementary Information). First, the comb in waveguide ``C'' is a three stage comb in contrast to primary--secondary combs observed in all other reported resonators to the best of our knowledge. The primary comb starts with two sidebands separated by $N\times FSR$, where N is integer. The secondary comb is still spaced by a multiple of resonator's FSR, whereas the third stage of comb formation fills the FSR--spaced resonator modes. Second, in all of the resonators shown in Fig. \ref{fig:waveguides} we observed discrete transitions between the modulation instability states of the primary comb that were not previously reported. The appearance of each new sideband pair is accompanied by reduction of N by unity and the overall growth of the primary comb span. These discrete steps eventually end with the formation of transitional comb states\cite{scalinglaws}, followed by a number of discrete transitions between stable and spectrally symmetric comb states. These final discrete transitions are also observed in resonator transmission in similarity to the steps observed on the nonlinear resonant curve during soliton formation \cite{solitonsMgF2, herrCrossing}. The discrete steps to the final states are observed in both two and three stage combs, whereas the three stage comb shown in Fig.\ref{fig:combs} is not symmetric. While behavior of our system is similar to the soliton formation regimes observed previously, further experiments are required to confirm such states in our system. 
\begin{table}[htbp]
{\caption{\textbf{Parameters of resonators and combs.} Quality factor of waveguide ``S'' was reduced from the critically coupled value of $400\times 10^6$ by overloading the output coupler. Loaded Q and cold coupling efficiency are the parameters measured at reduced pump power before and after comb generation. Intrinsic Q was derived from the linewidth measurements in critically coupled setting. }\label{tab:resonators}} 
\begin{center}
\begin{tabular}{|c|c|c|c|c|c|c|} \hline
Waveguide & Loaded Q (cold) & Intrinsic Q & Coupling (cold) \% & Coupling (critical) \%& N & FSR, GHz \\ \hline
S & $69\times 10^6$& $800\times 10^6$ & 38 &83& 29  & 46.06\\
A &$61\times 10^6$ &$69\times 10^6$& 21 &73& 29  & 45.96 \\
B & $36\times 10^6$ &$60\times 10^6$& 15 & 65 & 31  & 45.96 \\
C & $58\times 10^6$& $90\times 10^6$& 21 & 70 & 408 & 46.05\\
 \hline
\end{tabular}
\end{center}
\end{table}
Due to the single mode nature of micro--structured resonators the spectral signatures of mode crossing\cite{herrCrossing,grudininCrossing} were absent in all of the observed comb spectra with the exception of cavity ``S'' which still supports some higher order mode families.

\begin{figure}[htbp]
\centering
\includegraphics[width=12 cm]{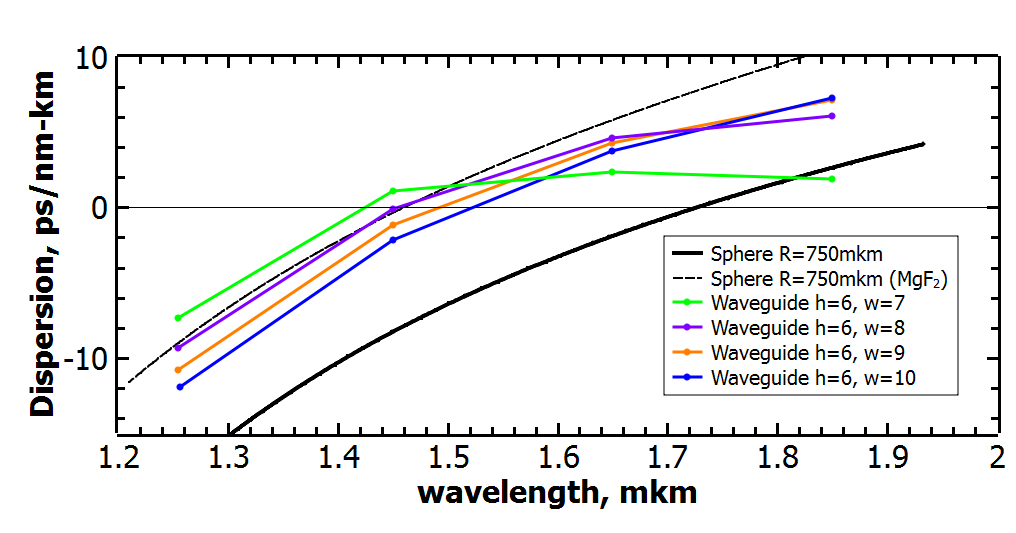}
\caption{\textbf{Computed dispersion of fluoride resonators with radius of 750 $\mu$m and varying height (h) and width (w) of the waveguide.} CaF$_2$ micro--structured resonator can have flat anomalous dispersion in telecom range.}
\label{fig:caf2}
\end{figure}

To further demonstrate the flexibility of dispersion engineering with respect to material choice, we show that it is possible to achieve anomalous and flattened GVD in a CaF$_2$ resonator at wavelength of 1550 nm. As shown in Fig. \ref{fig:caf2} the geometric dispersion of the waveguide brings the total dispersion into the anomalous regime. Indeed, our initial experimental results (Supplementary Information) confirm that changes in CaF$_2$ micro--structured waveguide geometry lead to comb generation regimes different from the non--micro--structured resonator\cite{grudinin-caf2}.

It is important to emphasize that in addition to flattened dispersion, these new micro--structured resonators support only one fundamental mode family of TE polarization. Higher order TE modes are not observable at least to the detector noise level. The quality factor of the fundamental TM mode family is at least 20 times lower for waveguides ``A--C'' compared to TE modes. This could be explained by the same geometric effect of mode leaking into the substrate that enables WGM single mode resonators\cite{singlemode, singlemodeQ} similar to resonator ``S'' where we also observed the difference in Q factors experimentally. However, we found that the difference in mode field distribution between TE and TM modes in our resonators is seemingly too small to account for such a Q difference. A more sophisticated theory of crystalline microresonators explaining this and other observed phenomena\cite{polarizationloss}  needs to be developed.

In summary we have presented a new resonator geometry amenable to dispersion engineering and spectrum control. We demonstrated the utility of dispersion engineering for efficient optical comb generation in microstructured crystalline resonators. The shape of the waveguide on the substrate provides design flexibilities in determining the set of modes, quality factors, and the geometric dispersion of the resonator. While we experimentally demonstrate that a flattened dispersion significantly improves the efficiency and span of combs generated in micro--structured resonators, a model linking the shape of the dispersion curve to the observed comb dynamics is yet to be developed. We note, however, that the observed comb broadening may be coincident with the change in third order dispersion sign. Moreover, both broadening and narrowing of the comb span were observed in our experiments (Supplementary Information). The nonlinear frequency comb generation process is rich and complex. Our investigation demonstrates the control needed in the crystalline resonator design for efficient broad comb generation and paves the way to direct octave--spanning dielectric micro--combs.\\

\noindent{\Large{\textbf{Methods}}}\\
{\small
To pump the modes of our resonators we used the Koheras Adjustik fiber laser with 1560 nm center wavelength, 20 kHz short--term linewidth and the IPG Photonics EDFA that provided power amplification to 300 mW. We used a computer controlled microfabrication technique developed in our lab to make the resonators. All of the resonators used in this work were fabricated from a z--cut (crystalline optical axis is parallel to the resonator symmetry axis) MgF$_2$ substrate and had the same diameter of 1.5 mm corresponding to FSR of 46 GHz. Optical Q factors were measured using a modulation sideband technique. The estimated mode area of the resonators is $A_m\simeq 25\times 10^{-12}m^2$. The angle polished SMF28e+ fiber couplers were used for evanescent excitation of modes and comb output. To record the spectra of the combs we used optical spectrum analyzers operating in the 650-2400 nm range (Yokogawa AQ6319 and AQ6375). Interestingly, the observed comb expanded beyond the nominal operating frequency range of the SMF-28 fiber used as an output channel.\\
\indent The FEM modelling of the fabricated axially symmetrical resonators requires knowledge of the waveguide cross section. For large WGM resonators this can be achieved by optical shadow microscopy (see Fig.\ref{fig:waveguides} - S). For micro--structured waveguides, SEM can be used in combination with ion milling that physically produces a cut across the waveguide. Obtaining a resonator cross section non--destructively remains a challenge, as SEM imaging requires gold coating to avoid charging effects. We have developed a technique that starts with making a resin cast of the resonator by direct contact and lift--off. The cast is then filled with a glue having different refractive index from the cast material. After hardening, the resulting replica can be cut and polished across the waveguide cross section resulting in images having sub--micrometer precision as shown in Fig.\ref{fig:waveguides}. While imperfections of this imaging technique can affect the results of numerical simulations, we found that resulting dispersion curves are robust against sub--micrometer noise in the waveguide boundary. Numerical modelling was used in this work as a design guideline.\\
\indent Since the real value of $D_2$ is around 10 kHz for our cavity size, we need to find the eigenfrequencies with the precision of around 1 kHz in order to achieve sufficient dispersion precision. To achieve this, we solve the three--dimensional vectorial Maxwell wave equation numerically, taking the axial symmetry into account \cite{oxborrow, sensor}.
We used a FreeFem++ package \cite{freefem}, with MUMPS solver implemented on a JPL supercomputer. Typically, the required eigenvalue precision is achieved with adaptive meshes containing up to 0.5 million elements. Second order GVD is obtained by computing the triplets $f_{l-1}, f_l, f_{l+1}$ for a number of wavelengths. For each $l$ a sequence of mesh refinements and iterative adjustments of the refractive indices with the Sellmeier equations is employed.
}

\noindent{\Large{\textbf{Acknowledgements}}}\\
The research described in this paper was carried out at the Jet Propulsion Laboratory, California Institute of Technology, under a contract with NASA. The authors thank the JPL supercomputing group and H. Lorenz--Wirzba for support; L. Baumgartel, D. Aveline, D. Strekalov and M. Wright for discussions and contributions to the experimental setup. \copyright 2014 California Institute of Technology. Government sponsorship acknowledged.

\noindent{\Large{\textbf{Author Contributions}}}\\
All authors contributed extensively to the work presented in this Letter. I.S.G and N.Y conceived the experiment, analyzed data, and participated in critical discussions. I.S.G designed and performed the experiment, and carried out numerical simulations. I.S.G. and N.Y. co-wrote the paper. N.Y. supervised the project. Authors declare no competing financial interests.

\end{document}